\documentclass[preprint,showpacs,preprintnumbers,amsmath,amssymb]{revtex4}

\usepackage{graphicx}% Include figure files
\usepackage{dcolumn}% Align table columns on decimal point
\usepackage{bm}% bold math

\usepackage{amssymb,bm,color}

\def\eqref#1{Eq.~(\ref{#1})}

\def\Eq#1{\begin{equation} #1 \end{equation}}
\def\Eqr#1{\begin{eqnarray} #1 \end{eqnarray}}
\def\Eqrsubl#1#2{\begin{subequations}\label{#1}\Eqr{#2}\end{subequations}}

\newcommand{\nn}{\nonumber}
\newcommand{\pd}{\partial}

\def\Xsp{{\rm X}}

\def\Ysp{{\rm Y}}

\def\X5sp{{\rm X}_5}
\def\Y3sp{{\rm Y}_3}
\def\Z3sp{{\rm Z}_3}

\def\lap{{\triangle}}
\def\e{{\rm e}}

\begin{document}

%\preprint{KU-TP }

\title{
Constraints on 
warped compactifications
%and de Sitter solutions
}% Force line breaks with \\

\author{Masato Minamitsuji}
\affiliation{
Department of Physics,
School of Science and Technology,
Kwansei Gakuin University, Sanda, Hyogo 669-1337, Japan.
}%

\author{Kunihito Uzawa}
\affiliation{%
Department of Physics, Kinki University,
Higashi-Osaka, Osaka 577-8502, Japan
}%

\date{\today}% It is always \today, today,

\begin{abstract}

We discuss 
a possible compactification of 
a higher-dimensional gravitational theory, 
which give rise to 
a de Sitter or an accelerating universe,
to the extent that it can  
describe a low energy limit of string theory.
Such analysis can be carried out by the usual 
integration over the internal space in the Einstein equations. 
The combined field equations in the higher-dimensional supergravity 
theory after integration
give
several new terms which may help to realize a de Sitter compactification.
After developing the general framework, 
we describe some specific examples 
involving the dilaton and mass parameter.  

\end{abstract}

\maketitle

%======================================%
%<<<<<<<<<<<<< SECTION 1 >>>>>>>>>>>>>>%
%======================================%

%T1>Introduction
\section{Introduction}
 \label{sec:introduction}

%{\color{red}
This paper is devoted to 
investigate the realizability of
non-singular de Sitter compactifications 
for higher-dimensional gravitational theories,
in particular in string theory. 
%Another motivation is to understand aspects of the 
Realization of cosmic acceleration is 
one of the requirements from observations
\cite{Riess:1998cb, Perlmutter:1998np, Riess:2001gk, Riess:2006fw, 
Kowalski:2008ez} (see also the review \cite{tsujikawa}
and references therein). 
At present, 
although
we have not successfully explained it yet 
from the point of view of string theory,
it would be desirable in future.

%One motivation for this study concerns
%higher-dimensional supergravity, which plausibly has at least some 
%sort of relationship to observational cosmology.
In that context,
cosmological model is obtained by mainly two approaches. 
One is an extension to a classical solution 
to the Einstein equation that was introduced in 
\cite{Gibbons:2005rt, Chen:2005jp}. Another is the 
construction of a lower-dimensional effective theory after integrating 
over the volume of internal space 
\cite{Duff:1986hr, Carroll:2001ih, Uzawa:2003ji, Uzawa:2003qh, 
Gibbons:2009dr}. 
The latter has a lot of 
works with respect to realizing a de Sitter universe. 
In the former case for type II supergravity and 
eleven-dimensional supergravity theory, 
there is the well-known obstacle about the NO-GO theorem
to obtain de Sitter compactifications \cite{Maldacena:2000mw}.
The NO-GO theorem forbids a de Sitter solution
in any higher dimensional theory 
which does not contain higher curvature corrections
and a nonpositive potential,
involves massless degrees of freedom with positive kinetic terms
and induces a finite effective gravitational constant.
Then an explicit proof has been made under the assumptions
that the warp factor is static,
and that the internal space is smooth and compact. 
There are several examples e.g., in
Ref. \cite{Emparan:2003gg, Townsend:2003fx,chen},
where time dependence (without the warp factor)
allows for accelerating universes.
%Therefore, 
de Sitter compactification solutions also
have been obtained by violating at least one of the above assumptions,
for example,
adding higher curvature corrections to the gravity action
(see e.g., \cite{mo0, mo,amo})
and 
introducing orbifold models with fluxes
(see e.g.,
\cite{Maldacena:2000mw, Danielsson:2009ff, Danielsson:2010bc, 
Caviezel:2009tu, Wrase:2010ew, Blaback:2010sj, Rosseel:2006fs}).
However, it has not been explicitly clarified that
de Sitter compactification solutions can be obtained
by allowing the time dependence of the warp factor
and without changing any other above assumption.
The main purpose of this work is 
to see
whether and how in this new class of compactifications
 the warped spacetime structure can realize 
de Sitter compactifications.

On the other hand,
in the recent years,
new development has been made 
in finding time dependent generalizations of 
$p$-brane solutions 
which are originally found by \cite{Gibbons:2005rt, Chen:2005jp}. 
%since the pioneering works \cite{Gibbons:2005rt, Chen:2005jp}.
Numerous number of time dependent solutions of D-branes and M-branes
like $p$-brane in any dimension have been obtained,
and their applications to cosmology
have been widely studied
in \cite{Binetruy:2007tu, Maeda:2009zi, Minamitsuji:2010kb}
(useful references are 
\cite{Kodama:2005fz, Kodama:2005cz}, and  
%partially 
physical explanations with some relevances to 
the present paper are 
\cite{Binetruy:2007tu, Maeda:2009tq, Maeda:2009zi, 
Maeda:2009ds, Maeda:2010yk, Maeda:2010ja, Minamitsuji:2010fp, 
Maeda:2010aj, Minamitsuji:2010kb, Minamitsuji:2010uz}).
A time dependent $p$-brane is explicitly calculable for 
any given link %in 
to a corresponding static solution. 
Indeed it was originally found 
with an explicit algebraic recipe 
(see \cite{Gibbons:2005rt, Chen:2005jp} for an accessible account).
Unfortunately, no $p$-brane solution which 
leads to a de Sitter or an accelerating expansion of our universe
has been found.

%except the 
%orientifold models with fluxes.
%Our analysis in this paper will shed light on how to make sense
%of this sort of compactifications, at least 
%in the context of gravity theory.
%Finally, 
%In the past few years, striking new results have 
%been obtained about the dynamical solution of $p$-brane 
%related to %the 
%string theory 

Let us review more on the properties of time 
dependent $p$-solutions and then 
deduce what should be revealed in this paper. 
In the case of the compactification in $p$-brane 
system discussed in \cite{Minamitsuji:2010kb}, 
the problem
is that of a decelerating expansion of the universe. 
%in the supergravity. 
When such a solution is found, 
the warp factor of the familiar static solutions 
is replaced by a function depending not only 
on the coordinates of the transverse space 
but on the linear function of time, studied in 
\cite{Binetruy:2007tu, Maeda:2009zi}.  
This might suggest generalizing our discussion 
to the case of intersecting branes as done in 
\cite{Maeda:2009zi,Minamitsuji:2010uz}. 
%We then go on in section \ref{sec:p} to 
%discuss the constraints of compactification
% of $p$-brane system. 
Hence there is a serious difficulty in obtaining an 
accelerating expansion
from these solutions 
because of a choice of ansatz of fields 
%in the higher-dimensional spacetime
with a given dilaton coupling parameter. 
Though these results are really natural in the viewpoint of the 
extension of the static solution,
it prevents us from obtaining an 
accelerating expansion unless we consider additional matter.
% on the brane. 
The previous %physics-based 
approaches %are 
have involved the exact time 
dependent solutions %of field equations 
with respect to the BPS $D$-brane solutions. 
For this purpose, we decide to %develop an approach to
investigate the realizability of
a de Sitter or an accelerating expansion 
through a more general warped compactification. 
%higher-dimensional theory. 
Thus, 
%Before going to explore any explicit %construction of 
%solution,
we will repeat the same argument 
of NO-GO theorem for warped compactificaton 
on the compact internal space,
but allowing time dependence in the warp factor.
%which relies on the 
%equation of motion for the warp factor. 
This is important to study how to choose the ansatz of fields 
and make use of the 
assumptions %for the background 
to get %the 
realistic cosmological %solutions in the 
warped compactificatons.

There are two issues to discuss cosmic 
acceleration in a warped compactification. 
As we mentioned above,
one is NO-GO theorem of warped compactification 
which is analyzed %by the field equations 
after integrating over
%the volume of 
the internal space,
which
does not make the exact solution manifest. 
The other is an approach which makes contact 
with %the 
exact solutions of higher-dimensional field equations. 
The description in this paper contains both aspects:
We discuss
the possibility of a four-dimensional accelerating universe 
from loopholes of NO-GO theorem. 
%are discussed.
We also obtain the exact solutions
which describe
de Sitter spacetime in warped compactifications. 
It would be highly desirable to discuss the de Sitter compactification 
from these two types of our calculation here because 
the condition of accelerating expansion as well as 
its computability are manifest.

We discuss the criterion for 
the possible de Sitter compactifications. 
%which will be determined by integrating 
%over the compact internal manifold.
We start with a compactification 
because of a single form field strength,
and then 
perform an integration over the compact manifold
so that it remains convergent. 
This gives the minimal framework
to analyze 
the possibilities of de Sitter compactification.
%We then 
An extension of our analyses to a more general theory
including a scalar field and multi- form field strengths
as in supergravity is straightforward.
We will see that a warped compactificaton leads to a new possibility  
of an accelerating expansion.

The organization of this paper is as follows.
In section \ref{sec:flux}, 
we consider the time dependent, warped
compactification that the internal 
space is smooth and compact.
%The background spacetime depends on time, 
%coming from the warp factor. 
The Einstein equations are used to %predict
derive some necessary condition
that de Sitter compactifications can be present, 
%under a certain condition for field strength. 
%In particular, 
for examples 
which were considered in \cite{Maldacena:2000mw} 
where the
%constraints of compactification have been worked out and 
no-go theorem of warped compactifications have been proposed.
%We show
%that these considerations agree with our constraints. 
We also look for the de Sitter compactifiaction with not only field strengths 
but also the 
scalar field.
%The fields support de Sitter compactification
%under the ansatz for fields appropriately. 
%In section \ref{sec:SUGRA}, 
We will focus on
the particular examples of supergravity theories,
the ten-dimensional type IIA theory 
%and $p$-branes
%as examples from string theory. 
as a example from string theory. 
%in \ref{sec:IIA}. 
In the case of IIA supergravity,
the Einstein equations imply that de Sitter compactifications may arise 
under certain conditions for the field strengths if the internal space 
is compact. Again, the results %that come 
from static compactifications 
can be compared %, in special cases, 
to results obtained in 
\cite{Maldacena:2000mw}.
There seems to be some room where 
the no-go theorem may be broken by a sort of 
time dependence from the warp factor. 
The last section \ref{sec:Discussions} 
 is devoted to summary and discussion.

%======================================%
%<<<<<<<<<<<<< SECTION 2 >>>>>>>>>>>>>>%
%======================================%

%T1>Compactifications
\section{Compactifications}
\label{sec:flux}

\subsection{Models with a single form field}
\label{subsec:flux}
We consider a gravitational theory with the metric $g_{MN}$
and an anti-symmetric tensor field strength of rank $n$. 
The action we consider is given by
\Eq{
S=\frac{1}{2\kappa^2}\int 
d^{D}x\sqrt{-g}
\left(R
 -\frac{1}{2\cdot n!}F^2
\right),
\label{D:action:Eq}
}
where $\kappa^2$ is the $D$-dimensional gravitational constant, 
$g$ is the determinant of the $D$-dimensional metric $g_{MN}$, 
and $F$ is an $n$-form field strength. 
After variations with respect to the metric, 
the $D$-dimensional Einstein equations are given by
\Eqr{
R_{MN}&=&T_{MN}-\frac{1}{D-2}g_{MN}T\nn\\
&=&\frac{1}{2\cdot n!}
\left(nF_{ML_2\cdots L_{n}} {F_N}^{L_2\cdots L_{n}}
-\frac{n-1}{D-2}g_{MN} F^2\right),
   \label{D:Einstein:Eq}
}
where $T$ is the trace of energy momentum tensor $T_{MN}$ 
which is defined by 
\Eq{
T_{MN}=\frac{1}{2\cdot n!}
\left(nF_{ML_2\cdots L_{n}} {F_N}^{L_2\cdots L_{n}}
-\frac{1}{2}g_{MN} F^2\right).
   \label{D:T:Eq}
}
Now we assume the $D$-dimensional metric in the form
\Eq{
ds^2=A^2(x, y)\left[q_{\mu\nu}(\Xsp)dx^{\mu}dx^{\nu}
  +u_{ij}(\Ysp)dy^idy^j\right],
 \label{D:metric:Eq}
}
where $q_{\mu\nu}(\Xsp)$ is a $d$-dimensional Einstein space 
metric which depends only on the $d$-dimensional coordinates $x^{\mu}$, 
and $u_{ij}(\Ysp)$ is the $(D-d)$-dimensional metric which
depends only on the $(D-d)$-dimensional coordinates $y^i$. 
Under the assumption \eqref{D:metric:Eq}, 
the $(\mu, \nu)$ component of Ricci tensor in \eqref{D:Einstein:Eq} 
gives
\Eqr{
&&R_{\mu\nu}=R_{\mu\nu}(\Xsp)-(D-2)\left(D_{\mu}D_{\nu}\ln A
-\pd_{\mu}\ln A\pd_{\nu}\ln A\right)\nn\\
&&~~~~-q_{\mu\nu}\left[\lap_{\Xsp}\ln A+\lap_{\Ysp}\ln A
+(D-2)\left(q^{\rho\sigma}\pd_{\rho}\ln A\pd_{\sigma}\ln A
+u^{ij}\pd_i\ln A\pd_j\ln A\right)\right],
    \label{D:Einstein-mn:Eq}
}
where $R_{\mu\nu}(\Xsp)$ is the Ricci tensor
of the metric $q_{\mu\nu}(\Xsp)$, and 
$\triangle_{\Xsp}$ and $\triangle_{\Ysp}$ are 
the Laplace operators on the space of 
$\Xsp$ %, and the space
and $\Ysp$, respectively.  
The trace of 
the $(\mu, \nu)$ components of Einstein equations %thus 
reduces to 
\Eq{
\frac{(D+d-2)}
{\alpha}A^{-\alpha}\lap_{\Xsp}A^{\alpha}
+\frac{d}{(D-2)A^{(D-2)}}
\lap_{\Ysp}A^{D-2}=R(\Xsp)+A^2\bar{T},
   \label{D:Einstein-mn2:Eq}
}
where $\alpha$ and $\bar{T}$ are defined by 
\Eqr{
\alpha&=&\frac{(D-2)(d-1)}{D+d-2},\\
    \label{D:alpha:Eq}
\bar{T}&=&-{T^{\mu}}_{\mu}+\frac{d}{D-2}T,
    \label{D:t:Eq}
}
respectively.
%In terms of 
Substituting Eq.~(\ref{D:T:Eq}) into Eq.~(\ref{D:t:Eq}),
% gives
we obtain
\Eq{
 \bar{T}=\frac{1}{2\cdot (n-1)!}\left[
 -F_{\mu M_1\cdots M_{n-1}}F^{\mu M_1\cdots M_{n-1}}
 +\frac{d(n-1)}{(D-2)n}F^2\right].
   \label{D:bar t:Eq}
}
The \eqref{D:bar t:Eq} implies $\bar{T}\ge 0$ if the field strength $F$ has 
only components along the $(D-d)$-dimensional space. 
In the case of the part of the field strength with components 
along the $d$-dimensional space, the trace over the index $\mu$ can be 
written by a particular order of contractions of the indices  
\cite{Maldacena:2000mw}
\Eq{
F_{\mu M_1\cdots M_{n-1}}F^{\mu M_1\cdots M_{n-1}}=\frac{d}{n}
%\beta 
F^2.
  \label{D:F:Eq}
}
Substituting the Eq.~(\ref{D:F:Eq}) into Eq.~(\ref{D:bar t:Eq}), we get
\Eq{
\bar{T}=
%-F^2\frac{d\left[\beta(D-2)-n+1\right]}{(D-2)n}\le 0.
-\frac{d(D-n-1)}{2(D-2)\cdot n!}F^2\ge 0,
     \label{D:trace T:Eq}
}
where %we used 
$F^2\le 0$ %because 
since the indices of the field strength are 
along the X space. 
%We consider the Eq.~(\ref{D:Einstein-mn2:Eq}).
Let us assume that the internal manifold should be compact. 
If the function $A$ depends only on the coordinate $y$,
after multiplying Eq.~(\ref{D:Einstein-mn2:Eq}) by $A^{D-2}$
and then integrating over the Y space,
the $y$-derivative term in the left-hand side does not contribute
since it becomes a total derivative term.
Thus, we obtain 
the consistency condition
\begin{eqnarray}
0= \frac{d}{(D-2)}
\int_{\Ysp} d^{D-d}y \sqrt{u}
\lap_{\Ysp}A^{D-2}
=\int_{\Ysp} d^{D-d}y \sqrt{u}
A^{D-2}\Big[R(\Xsp)+A^2\bar{T}\Big].
\end{eqnarray}
Nothing that $\bar{T}\geq 0$ for an $n$-form field
strength, 
we obtain $R(\Xsp)\leq 0$.
Therefore, de Sitter compactification can not be allowed
\cite{Maldacena:2000mw}.

We next consider the function $A$ which depends on the coordinate $x$
as well as on $y$. 
Then, we obtain
\begin{eqnarray}
\frac{(D+d-2)}{\alpha}
\int_{\Ysp} d^{D-d}y \sqrt{u}
A^{-\alpha+D-2}\lap_{\Xsp}A^{\alpha}
=\int_{\Ysp} d^{D-d}y \sqrt{u}
A^{D-2}\Big[R(\Xsp)+A^2\bar{T}\Big],
\end{eqnarray}
and thus,
we can not abandon the possibility $R(\Xsp)\ge 0$ if the function $A$ 
satisfies
\Eq{
\frac{(D+d-2)}{\alpha}A^{-\alpha+(D-2)}\lap_{\Xsp}A^{\alpha}>0.
  \label{D:A:Eq}
}
The Eq.~(\ref{D:A:Eq}) implies that there are de Sitter compactifications even 
if there is no higher derivative terms and is no scalar field 
which depends on time. 
%In fact, after multiplying Eq.~(\ref{D:Einstein-mn2:Eq}) 
%by $A^{(D-2)}$ and integrating over the compact internal space, 
%we get
%\Eqr{
%&&\frac{(D+d-2)}
%{\alpha}\left[A^{-(D-2)}D_{\rho}
%\left(A^{-\alpha+(D-2)}q^{\rho\sigma}\pd_{\sigma}A^{\alpha}\right)
%+\alpha\left\{\alpha-(D-2)\right\}
%q^{\rho\sigma}\pd_{\rho}\ln A\pd_{\sigma}\ln A\right]\nn\\
%&&%~~~-d_1(D-2)u^{ij}\pd_i\ln A\pd_j\ln A
%=R(\Xsp)+A^2\bar{T},
%    \label{D:A-2:Eq}
%}
%where we note 
%\Eq{
%\alpha-%2
%(D-2)=-(D-2)%\left[2(D-2)+d_1+1\right]
%(D-1)(D+d-2)^{-1}<0,~~~~
%{\rm for}~~~D>2.
%}
%For $D>2$, recalling that $q^{\rho\sigma}$ become negative for the time
%component,
%the kinetic term of $A$ is 
%the positive contribution to the Ricci scalar of the four-dimensional 
%universe. 
%Hence, the time dependence of the warp 
%factor supports the de Sitter compactifications. 

In principle we could have functions of scalar fields multiplying 
these expressions, as we have in some supergravity theories, 
and we could also have many types of $n$-form fields,
as we will discuss in the next subsection. 
%We will not indicate these explicitly but it is obvious how to extend
%the following arguments to those cases. 
In fact, the action (\ref{D:action:Eq}) is a straightforward 
generalization of the case of eleven-dimensional supergravity 
or ten-dimensional supergravity without dilaton coupling.

If there is 
%In 
a hypersurface %region 
where $A=0$, 
%there  
a curvature singularity appears there.
%at the hypersurface of $A=0$.
% where $A$ vanishes. 
%Hence we can not discuss the 
%compactification at $A=0$. 
%If 
Even in such a case,
we can always consider the region $A>\epsilon$ 
where the singularities are left out for small $\epsilon$,
and then apply 
%we find 
the same arguments for the integral %\eqref{D:integral:Eq} 
(see \cite{Maldacena:2000mw} for details).
%Thus it is possible to find de Sitter 
%compactification if function $A$ depends on $x^{\mu}$.

%We have discussed the condition for the $d$-dimensional space 
%via the Einstein equations. The compactifications under our argument is 
%valid only for $d$-dimensional spacetime locally because 
%Eq.~(\ref{D:A:Eq}) yields after the integrating Eq.~(\ref{D:Einstein-mn2:Eq})
% over the internal space.
%The Ricci scalar of $d$-dimensional spacetime in the $D$-dimensional 
%theory will be determined by the Eq.~(\ref{D:Einstein-mn2:Eq}) after 
%integration of the $d$-dimensional coordinates $x$ as well as the 
%internal coordinates $y$.  
%If Eq.~(\ref{D:Einstein-mn2:Eq}) 
%integrated by the coordinates $x$ and $y$, the kinetic term of $A$ 
%contributes positive sign in the left hand side. 
%Then, it is possible to find de Sitter compactification if the 
%function $A$ depends on time. 
%This is because 
%the ansatz of the metric (\ref{D:metric:Eq}) is rather restricted. 
%We will also find de Sitter compactification in sec.~\ref{sec:p}, 
%if we choose another metric ansatz. 

Let us go back to the Einstein equations.
($\mu, i$) and ($i, j$) components 
of Einstein equations are given by
\Eqrsubl{D:Ein:Eq}{
&&(D-2)A\pd_{\mu}\pd_iA^{-1}=\frac{1}{2\cdot (n-1)!}F_{\mu A\cdots B}
{F_i}^{A\cdots B},
   \label{D:Ein-mi:Eq}\\
&&(D-d)\left[A^{-1}\lap_{\Xsp}A
+(D-3)q^{\rho\sigma}\pd_{\rho}\ln A\pd_{\sigma}\ln A\right]
+(2D-d-2)A^{-1}\lap_{\Ysp}A\nn\\
&&~~~+\left[(D-d)(D-3)-2(D-2)\right]u^{ij}\pd_i\ln A\pd_j\ln A
=R(\Ysp)+A^2\tilde{T},
  \label{D:Ein-ij:Eq}
}
where $\tilde{T}$ is defined by
\Eq{
\tilde{T}\equiv -{T^i}_i+\frac{(D-d)}{D-2}T.
    \label{D:t2:Eq}
}
Eq.~(\ref{D:t2:Eq}) gives $\tilde{T}<0$ if the field strength $F$ has 
only components along the $(D-d)$-dimensional space. 
In the case of the part of the field strength with components 
along the $d$-dimensional space, we get again $\tilde{T}<0$ due to 
Eq.~(\ref{D:F:Eq}). 

Using Eqs.~(\ref{D:Einstein-mn2:Eq}) and (\ref{D:Ein-ij:Eq}), we get
\Eqr{
&&(D-1)%(D-2)
\left[(D-d-2)\lap_{\Xsp}\ln A
-(D-2)q^{\rho\sigma}\pd_{\rho}\ln A\pd_{\sigma}\ln A
-d \lap_{\Ysp}\ln A\right]\nn\\
&&~~~~=%(D-2)
%\left[
(D-d-1)R(\Xsp)-dR(\Ysp)%\right]
+A^2%(D-2)
\left[(D-d-1)\bar{T}
-d\tilde{T}\right].
   \label{D:Ein2:Eq}
}
The second term of the left hand side of Eq. (\ref{D:Ein2:Eq}),
i.e., the kinetic term of $A$,
contributes positively to it. Therefore,
the de Sitter compactification may be allowed for  
the case of $\lap_{\Xsp}\ln A>0$ and $R(\Ysp)>0$ because 
of $(D-d-1)\bar{T}-d\tilde{T} \ge 0$ for $D>d+1$.

Before closing this subsection,
some remarks are in order. 
In getting the four-dimensional
cosmology from Eq. (\ref{D:metric:Eq}),
there would be two approaches.
The first one is integrating over the $\Ysp$ space,
giving rise to the four-dimensional effective theory with 
the de Sitter universe.
The second one is
assuming that we are living at some particular place
of the internal space, like a braneworld picture.
Then, the effective four dimensional metric 
$A^2(x) q_{\mu\nu}(\Xsp)dx^{\mu}dx^{\nu}$
may be no longer de Sitter universe,
even being decelerating universe.

%We will comment about the condition
Our previous argument is applied
not only to the cases that the $d$-dimensional spacetime $\Xsp$ is 
the Einstein space,
but also to those of more general $d$-dimensional metric.
%Let us
For instance,
let us consider the case that the $d$-dimensional metric is given by 
the Robertson-Walker form
\Eq{
ds^2(\Xsp)=-dt^2+t^{2\lambda}\delta_{ab}dx^adx^b,
}
where $\lambda$ is the parameter, and 
$\delta_{ab}$ denotes the $(d-1)$-dimensional 
Euclidean space. The Ricci scalar $R(\Xsp)$ is expressed as  
\Eq{
R(\Xsp)=d(d-1)\lambda\left(\lambda-\frac{2}{d}\right)t^{-2}\,.
}
Then, the Ricci scalar of the X space becomes $R(\Xsp)> 0$  
only if the parameter $\lambda$ satisfies $\lambda> 2/d$. 
Hence, for $2/d< \lambda\le 1$, the $d$-dimensional spacetime 
is not accelerating expansion while the Ricci scalar of the 
X space is positive. 
For $d=4$, a positive $R(\Xsp)$ corresponds to
an universe expanding faster than the radiation dominated universe
with $\lambda=\frac{1}{2}$. 
The above thing will also be true also for the discussions
in the rest of this paper,
although we mainly focus on the case that 
$\Xsp$ is an Einstein space.

\subsection{Models with a scalar field and multi- form fields}

We now %look for 
discuss the de Sitter compactification 
in the theory
with the 
multi- form field strengths as well as the scalar field,
which is more relevant for the supergravity theories. 
%We will see that de Sitter compactification will 
%be emerged if the ansatz of fields are taken appropriately. 

We consider a gravitational theory with the metric $g_{MN}$,
dilaton $\phi$, the cosmological constant $\Lambda$, 
and anti-symmetric tensor fields of rank $n_I$. 
The action %we consider 
in the Einstein frame is given by
\Eq{
S=\frac{1}{2\kappa^2}\int d^Dx\sqrt{-g}\left(R-2\e^{\chi\phi}
\Lambda
 -\frac{1}{2}g^{MN}\pd_M\phi \pd_N\phi
 -\sum_{I}\frac{1}{2\cdot n_I!}\e^{c_I\phi}F_{n_I}^2\right),
\label{f:action:Eq}
}
where $\kappa^2$ is the $D$-dimensional gravitational constant, 
$g$ is the determinant of the $D$-dimensional metric $g_{MN}$, 
$F_{n_I}$ is $n_I$-form field strength, 
and $c_I$, $\chi$ are constants. 
%After variations with respect to the metric, and the dilaton,  
%we obtain the field equations,
The field equations are given by
\Eqrsubl{f:field equations:Eq}{
&&\hspace{-1cm}R_{MN}=\frac{2}{D-2}\e^{\chi\phi}\Lambda g_{MN}
+\frac{1}{2}\pd_M\phi \pd_N \phi\nn\\
&&~~~~~~+\sum_I\frac{1}{2\cdot n_I!}\e^{c_I\phi}
\left(nF_{MA_2\cdots A_{n_I}} {F_N}^{A_2\cdots A_{n_I}}
-\frac{n-1}{D-2}g_{MN} F_{n_I}^2\right),
   \label{f:Einstein:Eq}\\
&&\hspace{-1cm}\lap\phi=\sum_I
\frac{c_I}{2\cdot n_I!}\e^{c_I\phi}F^2_{n_I}
+2\chi \e^{\chi\phi}\Lambda,
   \label{f:scalar:Eq}
}
where $\lap$ is the $D$-dimensional Laplace operator. 
%, and 
%$\nabla_M$ is the covariant derivative with respect to the 
%metric $g_{MN}$ and $\lap:=\nabla^M \nabla_M $. 

%We now assume that the $D$-dimensional metric takes the form
We assume the form of the $D$-dimensional metric as
\Eqr{
ds^2=A^2(x, y)q_{\mu\nu}(\Xsp)dx^{\mu}dx^{\nu}
+B^2(x, y)u_{ij}(\Ysp)dy^i\,dy^j,
   \label{f:metric:Eq}
}
where $q_{\mu\nu}(\Xsp)$ is a $d$-dimensional Einstein space 
metric which depends only on the $d$-dimensional coordinates $x^{\mu}$, 
and $u_{ij}(\Ysp)$ is the $(D-d)$-dimensional metric which
depends only on the $(D-d)$-dimensional coordinates $y^i$. 
%In terms of
With the same procedure as section \ref{sec:flux}-A,
%and \ref{sec:IIA}, 
the scalar field equation and 
the trace of 
the $(\mu, \nu)$ components of the Einstein equation are given by 
\Eqrsubl{f:equations:Eq}{
&&A^{-d}B^{-(D-d)}D_{\mu}\left(A^{d-2}B^{D-d}
q^{\mu\nu}\pd_{\nu}\phi\right)
+A^{-d}B^{-(D-d)}D_i\left(A^{d}B^{D-d-2}u^{ij}\pd_j\phi\right)
\nn\\
&&~~~~=2\chi \e^{\chi\phi}\Lambda
 +\frac{1}{2}\sum_I\frac{c_I}{n_I!}\e^{c_I\phi}
 \left[\theta(n_I-d)F_{n_I\,{\rm ex}}^2+F_{n_I\,{\rm in}}^2\right],
\label{f:s-equation:Eq}\\
&&2(d-1)A^{-1-d/2}D_{\mu}\left(A^{-2+d/2}q^{\mu\nu}\pd_{\nu}A\right)
+(D-d)A^{-d}B^{-1}D_{\mu}
\left(A^{d-2}B^{-1}q^{\mu\nu}\pd_{\nu}B\right)\nn\\
&&~~~~~+dA^{-d}B^{-(D-d)}D_i\left(A^{d-1}B^{D-d-2}u^{ij}\pd_jA\right)
=A^{-2}R(\Xsp)-\frac{2\Lambda d}{D-2}\e^{\chi\phi}
-\frac{1}{2}g^{\mu\nu}\pd_{\mu}\phi\pd_{\nu}\phi\nn\\
&&~~~~~-\frac{d}{2(D-2)}\sum_I
\frac{1}{n_I!}\e^{c_I\phi}\left[(D-n_I-1)\theta(n_I-d)
F_{n_I\,{\rm ex}}^2-(n_I-1)F_{n_I\,{\rm in}}^2\right],
   \label{f:Einstein munu:Eq}
}
where $\theta(a)$ is defined by %(\ref{m:theta:Eq}), 
\Eqr{
\theta(a)&=&\left\{
\begin{array}{cc}
 1&~{\rm if}~~a\ge 0 \\
 0&~~{\rm if}~~a<0\,.
\end{array} \right.
 \label{m:theta:Eq}
   }
Here $F^2_{n_I\,{\rm in}}$ is the square of 
a tensor with components purely in the
internal dimensions and $F^2_{n_I\,{\rm ex}}$ 
is %the square of the tensor
that with components 
along the $d$-dimensional spacetime directions.
$D_{\mu}$, $D_i$ are the covariant derivatives with respect to
the metric $q_{\mu\nu}$, $u_{ij}$, respectively.
After some algebra, these two equations can be combined into
\Eqr{
&&\hspace{-0.5cm}
dD_{\mu}\left(A^{d-2}B^{D-d}q^{\mu\nu}\pd_{\nu}\phi\right)
+dD_i\left(A^{d}B^{D-d-2}u^{ij}\pd_j\phi\right)
+d(D-2)D_i\left(A^{d-1}B^{D-d-2}u^{ij}\pd_jA\right)\nn\\
&&~~~\hspace{-0.3cm}+2(d-1)(D-2)A^{-1+d/2}B^{D-d}
D_{\mu}\left(A^{-2+d/2}q^{\mu\nu}\pd_{\nu}A\right)\nn\\
&&~~~\hspace{-0.3cm}+(D-d)(D-2)B^{D-d-1}D_{\mu}
\left(A^{d-2}B^{-1}q^{\mu\nu}\pd_{\nu}B\right)\nn\\
&&~~~\hspace{-0.3cm}=(D-2)A^{d-2}B^{D-d}R(\Xsp)
+2\Lambda d(\chi-1)\e^{\chi\phi}
-\frac{1}{2}(D-2)A^{d}B^{D-d}g^{\mu\nu}\pd_{\mu}\phi\pd_{\nu}\phi\nn\\
&&~~~\hspace{-0.3cm}+\frac{d}{2}A^{d}B^{D-d}
\sum_I\frac{\e^{c_I\phi}}{n_I!}
\left[-\left(D-n_I-1-c_I\right)\theta(n_I-d)F_{n_I\,{\rm ex}}^2
+\left(c_I+n_I-1\right)F_{n_I\,{\rm in}}^2\right].
  \label{f:field eq:Eq} 
}
If the functions $A$ and $B$ depend only on $y^i$, the right hand side of 
\eqref{f:field eq:Eq} becomes positive %for even $D$ and $d$ 
because the square of a tensor 
with components in the internal dimensions is positive and the 
square of the tensor with components 
along the $d$-dimensional spacetime directions is negative. 
If we have a compact internal manifold, 
the left hand side in \eqref{f:field eq:Eq} is zero since 
we have a total derivative after we integrate \eqref{f:field eq:Eq} 
over the manifold. 
On the other hand, the right hand side in \eqref{f:field eq:Eq} is non-zero 
unless  $R(\Xsp)=0$ and $F_{n_I}=0$. Hence, for a compact internal manifold, 
there are no nonsingular de Sitter compactifications. 
%to either de Sitter or Minkowski space.
However, if $\Lambda(\chi-1)<0$ 
%or $A^{d-2}B^{D-d}<0$, 
%if $D$ and $d$ are odd numbers, the function $A^{d}B^{D-d}$ 
%can take a negative value.  It is possible to find 
%de Sitter compactification even if the functions $A$ and $B$ depend only 
%on $y^i$. 
it %is 
may be
possible to find de Sitter compactification even if the functions 
$A$ and $B$ depend only on $y^i$.

Furthermore, if the functions $A$ and $B$ depend on both $x^{\mu}$ and $y^i$, 
the left hand side in \eqref{f:field eq:Eq} is not zero due to the 
contribution of the terms $\pd_{\mu}A$ and $\pd_{\mu}B$. 
Then, we can find the nonsingular de Sitter compactifications. 
%to either de Sitter or Minkowski space.

%In order to discuss the compactification which is not locally valid in 
%$d$-dimensional spacetime, we have to integrate \eqref{f:field eq:Eq} 
%over not only X but Y. 
%the sign of the Ricci scalar $R(\Xsp)$ will be able to find  
%after the integration over X. 
%Integrating \eqref{f:field eq:Eq} over X space by parts we conclude that 
%the left hand side in \eqref{f:field eq:Eq} may allow $R(\Xsp)\ge 0$ 
%even if there is no scalar field in the background. 

%======================================%
%<<<<<<<<<<<<< SECTION 3 >>>>>>>>>>>>>>%
%======================================%
%T1>Applications to supergravities
%\section{Applications %to supergravities
%}
%\label{sec:SUGRA}
%In this section,
%we discuss the possibility of de Sitter compactifications 
%of supergravity theories by use of the discussion 
%in the previous section. 

%It turns out that de Sitter compactifications 
%is possible for the ten-dimensional massive IIA supergravity model 
%if the warp factor depends on time. 
%In what follows, we present such a compactification explicitly.

%%%%%%%%%%%%%%%%%%%%%%%%%%%%%%%%%%%%%%%%%%%%%%%%%%%%%%%%

%T2>IIA compactifications
\subsection{de Sitter compactifications in the type IIA theory}
\label{sec:IIA}

In this subsection, 
we discuss the possible de Sitter compactifications 
of the ten-dimensional massive IIA supergravity.
The action for the massive IIA supergravity in the Einstein 
frame can be written as \cite{Lust:2004ig}
\Eqr{
S&=&\frac{1}{2\kappa^2}\int d^{10}x\sqrt{-g} 
\left(R-\frac{1}{2}g^{MN}\pd_M\phi\pd_N\phi
 -\frac{1}{2\cdot 2!}\e^{3\phi/2}F^2
 -\frac{1}{2\cdot 3!}\e^{-\phi}H^2\right.\nn\\
 &&\left.-\frac{1}{2\cdot 4!}\e^{3\phi/2}G^2
 -\frac{1}{2}\e^{5\phi/2}m^2\right)
 +S_{\rm CS},
\label{m:action:Eq}
}
where $S_{\rm CS}$ denotes the action for Chern-Simons term, 
$R$ is the ten-dimensional 
Ricci scalar with respect to the ten-dimensional metric 
$g_{MN}$, $\kappa^2$ is the ten-dimensional 
gravitational constant, $m$ is constant, $g$ is the determinant  
of the ten-dimensional metric $g_{MN}$, and $F$, $H$, $G$ are 
2-form, 3-form, 4-form field strengths, respectively.
The expectation values of fermionic fields are assumed to be zero.

%After variations with respect to the metric and dilaton, the Einstein 
%equations and the field equations are given by
The ten-dimensional action (\ref{m:action:Eq}) gives following 
field equations:
\Eqrsubl{m:equations:Eq}{
&&\hspace{-0.4cm}
R_{MN}=\frac{1}{2}\pd_M\phi \pd_N \phi+\frac{1}{16}m^2\e^{5\phi/2}g_{MN}
+\frac{1}{2\cdot 2!}\e^{3\phi/2} 
\left(2F_{MA} {F_N}^{A}-\frac{1}{8}g_{MN} F^2\right)\nn\\
&&\hspace{1cm}+\frac{1}{2\cdot 3!}\e^{-\phi} 
\left(3H_{MAB} {H_N}^{AB}
-\frac{1}{4}g_{MN} H^2\right)\nn\\
&&\hspace{1cm}+\frac{1}{2\cdot 4!}\e^{\phi/2} 
\left(4G_{MABC} {G_N}^{ABC}
-\frac{3}{8}g_{MN} G^2\right),
   \label{m:Einstein:Eq}\\
&&\hspace{-0.5cm}
\lap\phi=\frac{5}{4}\e^{5\phi/2}m^2+\frac{3}{4\cdot 2!}
\e^{3\phi/2}F^2-\frac{1}{2\cdot 3!}\e^{-\phi}H^2
+\frac{1}{4\cdot 4!}\e^{\phi/2}G^2,
   \label{m:scalar:Eq}
}
where $\lap$ is the Laplace operator with respect to the metric 
$g_{MN}$. 

Now the ten-dimensional metric is assumed to be \eqref{D:metric:Eq}. 
Under the same procedure as in sec.~\ref{sec:flux},
the trace of 
the $(\mu, \nu)$ components of the Einstein equations is given by 
%\Eqrsubl{m:cEinstein:Eq}{
\Eqr{
&&\frac{(d+8)}{\alpha}A^{-\alpha-2}\lap_{\Xsp}A^{\alpha}+\frac{d}{8}A^{-10}
\lap_{\Ysp}A^8=A^{-2}R(\Xsp)-\frac{1}{2}g^{\mu\nu}\pd_{\mu}\phi\pd_{\nu}\phi
\nn\\
&&~~~~+\frac{d}{16}\left(
\frac{1}{2!}\e^{3\phi/2}F_{\rm in}^2+\frac{2}{3!}\e^{-\phi/2}H_{\rm in}^2
+\frac{3}{4!}\e^{\phi/2}G_{\rm in}^2\right)-\frac{d}{16}\e^{5\phi/2}m^2\nn\\
&&~~~~-\frac{d}{16}\left[\frac{7}{2!}\theta(2-d)
\e^{3\phi/2}F_{\rm ex}^2
+\frac{6}{3!}\theta(3-d)\e^{-\phi}H_{\rm ex}^2
+\frac{5}{4!}\theta(4-d)\e^{\phi/2}G_{\rm ex}^2\right],
  \label{m:cEinstein-mn:Eq}
%&&16A\pd_{\mu}\pd_iA^{-1}=\pd_{\mu}\phi\pd_i\phi,\nn\\
%  \label{m:cEinstein-mi:Eq}
}
where $\triangle_{\Xsp}$ and $\triangle_{\Ysp}$ are 
the Laplace operators on the space of 
${\rm \Xsp}$ and the space ${\rm \Ysp}$, and 
$R_{\mu\nu}(\Xsp)$ is the Ricci tensor of the metrics 
$q_{\mu\nu}$, and the function $\theta(a)$ is defined by 
Eq. (\ref{m:theta:Eq}), and $\alpha$ is defined by \eqref{D:alpha:Eq}, and 
$F_{\rm in}$, $H_{\rm in}$, $G_{\rm in}$ are the squares of 
a tensors with components purely in the
internal dimensions and $F_{\rm ex}$, $H_{\rm ex}$, $G_{\rm ex}$ 
are the squares of the tensors with components 
along the $d$-dimensional spacetime directions. 
%, and $\beta$ is defined by
%\Eq{
%\beta=\frac{8(9-d_1)}{18-d_1}.
%} 
The components $F_{\rm ex}$, $H_{\rm ex}$, $G_{\rm ex}$ can only 
appear if the rank of the tensor is bigger or equal to $d$.

Substituting \eqref{D:metric:Eq}  
into the equation of motion for the scalar field  
\eqref{m:scalar:Eq}, we find 
\Eqr{
&&A^{-10}\left[D_{\mu}\left(A^8q^{\mu\nu}\pd_{\nu}\phi\right)
+D_i\left(A^8u^{ij}\pd_j\phi\right)\right]\nn\\
&&~~~~=\frac{5}{4}\e^{5\phi/2}m^2+\frac{1}{2}\left(
\frac{3}{2\cdot 2!}\e^{3\phi/2}F^2
-\frac{1}{3!}\e^{-\phi}H^2+\frac{1}{2\cdot 4!}\e^{\phi/2}G^2\right),
   \label{m:scalar2:Eq}
}
where $D_{\mu}$, $D_i$ are the covariant derivatives with respective to 
the metric $q_{\mu\nu}$, $u_{ij}$. 
In terms of \eqref{m:cEinstein-mn:Eq} and \eqref{m:scalar2:Eq}, 
we get
\Eqr{
&&\frac{10(d+8)}{\alpha}
A^{-\alpha+8}\lap_{\Xsp}A^{\alpha}+\frac{5d}{4}\lap_{\Ysp}A^8
+\frac{d}{2}D_{\mu}\left(A^8q^{\mu\nu}\pd_{\nu}\phi\right)
+\frac{d}{2}D_{i}\left(A^8u^{ij}\pd_j\phi\right)\nn\\
&&~~~~=10A^8R(\Xsp)-5A^{10}g^{\mu\nu}\pd_{\mu}\phi\pd_{\nu}\phi
+dA^{10}\left(\frac{1}{2!}\e^{3\phi/2}F_{\rm in}^2
+\frac{1}{3!}\e^{-\phi}H_{\rm in}^2
+\frac{2}{4!}\e^{\phi/2}G_{\rm in}^2\right)\nn\\
&&~~~~~~~-dA^{10}\left[\frac{4}{2!}\theta(2-d)\e^{3\phi/2}F_{\rm ex}^2
+\frac{4}{3!}\theta(3-d)\e^{-\phi}H_{\rm ex}^2
+\frac{3}{4!}\theta(4-d)\e^{\phi/2}G_{\rm ex}^2
\right].
    \label{m:field eq:Eq}
}
%where $\theta(a)$ is defined by Eq. (\ref{m:theta:Eq}).
If the function $A$ depends only on $y^i$, the right hand side of 
\eqref{m:field eq:Eq} is positive since the square of a tensor 
with components in the internal dimensions is positive and the 
square of the tensor with components 
along the $d$-dimensional spacetime directions is negative. 
For a compact internal manifold, 
the left hand side in \eqref{m:field eq:Eq} becomes zero since 
we have a total derivative after we integrate \eqref{m:field eq:Eq} 
over the internal space. 
On the other hand, the right hand side in \eqref{m:field eq:Eq} is non-zero 
unless  $R(\Xsp)=0$ and $F=H=G=0$. Hence, 
there are no nonsingular de Sitter compactifications. 
%The $d$-dimensional space is Minkowski space only if we can be turned
%on $F=H=G=0$. 
%to either de Sitter or Minkowski space.
However, if the function $A$ depends not only on $y^i$ but on $x^{\mu}$, 
the left hand side in \eqref{m:field eq:Eq} is not zero due to the 
contribution of $A^{-\alpha+8}\lap_{\Xsp}A^{\alpha}$. 
Then, it is possible to find the de Sitter compactifications. 
%to either de Sitter or Minkowski space. 

%As we have discussed in sec.\ref{sec:flux}, the Eq.~(\ref{m:field eq:Eq}) 
%integrated by the coordinates $y$ gives the compactifications for 
%$d$-dimensional spacetime locally. 
%After integration of the $d$-dimensional coordinates 
%$x$ as well as the internal coordinates $y$ 
%in the Eq.~(\ref{m:field eq:Eq}), 
%the first term of the left hand side allows $R(\Xsp)\ge 0$.
%Then, we can find de Sitter compactification due to the 
%kinetic term of the function $A$.

%If we integrate \eqref{m:field eq:Eq} over X space, 
%the first term of the left hand side allows $R(\Xsp)\ge 0$. 

If we consider the regions where the function $A$ vanishes 
the curvature singularity will again emerge. 
%The warp factor 
%in general is related to the scalar field in the supergravity, 
%the scalar field diverges at $A=0$. 
Then we can not analyze the 
construction of compactifications any further. 
If we consider again the region where the curvature singularities 
can be left out, we can find the de Sitter compactification 
by the same steps as in the section \ref{sec:flux}. 

%We next consider the \eqref{m:cEinstein-ij:Eq}. By using Eqs.~
%(\ref{m:cEinstein-ij:Eq}) and (\ref{m:scalar2:Eq}), we get 

%======================================%
%<<<<<<<<<<<<< SECTION 4 >>>>>>>>>>>>>>%
%======================================%
%T1>Discussions
\section{Discussions}
  \label{sec:Discussions}

In this work, we proposed a relatively simple and explicit 
class of de Sitter compactification in gravity theory coupled to form fields,
including the supergravity theories. 
We showed how a few ingredients
suffice to produce several new terms 
which exhibit de Sitter compactifications.
%involved in the basic matter  fields.
 In sections \ref{sec:flux},  
%and \ref{sec:SUGRA}, 
we have noted that the condition of the de Sitter compactification 
can be interpreted formally as %saying  the one that 
the condition 
that the warp factor determines a compactification.
%Thus, it is sometimes called the warped compactifications.
In the warped compactification, this 
interpretation is somewhat formal because the contribution of 
the warp factor can strictly determine the lower-dimensional geometry. 
Especially %in the case of time dependent space, 
to have 
terms from the warp factor 
leads to the possibility of the de Sitter compactifications.

%Some progresses toward a general framework for de Sitter spacetime 
%have appeared in \cite{Minamitsuji:2010uz}. 
The time dependence of the warp factor may also 
%support de Sitter compactifications is somewhat analogous to the fact 
support to obtain more general accelerating universes %of Universe 
since our analysis can be applied 
to the cases that the four-dimensional geometry
is not an Einstein space, as mentioned in Sec. II-A.
It is also worth mentioned that
the time dependence of the warp factor plays important roles 
in other dynamical systems,
for example, it gives the dynamical black hole solutions
in the asymptotically de Sitter spacetime  
in the Einstein-Maxwell theory \cite{Maki:1992tq, Minamitsuji:2010uz}.

Clearly, an important direction for further work is fleshing 
out further the methods in section \ref{sec:flux} for 
finding the cosmological model from the exact solution of Einstein equations. 
A convenient feature of the background is its simple ansatz for matter 
fields and metric in higher-dimensional spacetime, which make a controlled 
analysis of the warp factor and the internal 
space possible. In the analysis of the four-dimensional effective theory, 
the curvature of the internal and KK 5-brane configuration facilitates
the de Sitter compactification by introducing useful competing forces 
\cite{Silverstein:2007ac}.

%The Einstein equations give a constraint on compactifications. 

One of the consideration is that the metric flux and KK 5-branes yield 
the de Sitter spacetime in the four-dimensional effective theory 
\cite{Silverstein:2007ac, Villadoro:2007tb, Villadoro:2007yq}. 
Since these matter fields couple to the moduli of the internal space, 
the effects of fields modify the potential
to realize the four-dimensional de Sitter spacetime. 
It would be interesting to apply our construction to 
the problems of explicitly modeling inflation in string theory. 
One question is whether the time dependence of the warp factor  
could help to tune the inflationary scenario and accelerating expansion 
of our Universe. 
It might also be interesting to introduce higher derivative corrections 
or source term corresponding to the field strengths to these models, 
perhaps using an orientifold plane within the bulwark of D-branes and 
NS-branes to form brane constructions of the relevant field theories.
Some rudimentary model-building observations based 
on this mechanism for obtaining the de Sitter compactification 
will appear elsewhere.

%T1>Acknowledgments
\section*{Acknowledgments}
K.U. would like to thank H. Kodama, M. Sasaki, N. Ohta, and T. Okamura
for continuing encouragement. K.U. is supported by Grant-in-Aid for 
Young Scientists (B) of JSPS Research, under Contract No. 20740147.

%======================================%
%<<<<<<<<<<<<< REFERENCE >>>>>>>>>>>>>>%
%======================================%

%T1>References

\end{document}